\documentclass{article}
\usepackage{graphicx}
\def\be{\begin{equation}}
\def\ee{\end{equation}}
\begin{document}

\title{Personal recollections:
Frascati and the search for gravitational waves at the \it Istituto Nazionale di Fisica Nucleare \rm (INFN).
}
\author{
G.Pizzella\\
INFN Laboratori Nazionali di Frascati\\
Email:guido.pizzella@lnf.infn.it}

\maketitle
\date{}

\section{Introduction}

Along the way named Via Isacco Newton in the \it Laboratori Nazionali di Frascati \rm  (LNF) of the \it Istituto Italiano di Fisica Nucleare \rm (INFN) in Piazza Albert Einstein  one meets a building with the inscription NAUTILUS. In this building a cryogenic detector of gravitational waves is installed, the most sensitive one in the world at the end of the 90s. 

NAUTILUS started to operate in 1991 and will be turned off in June 2016.  I think interesting to remember briefly how it has come to carry out this research in the Frascati National Laboratories.

The story begins in 1961, when Edoardo Amaldi attended in Varenna lectures on  gravitational  waves by Joe Weber.
\begin {figure}
\includegraphics[width=130mm]{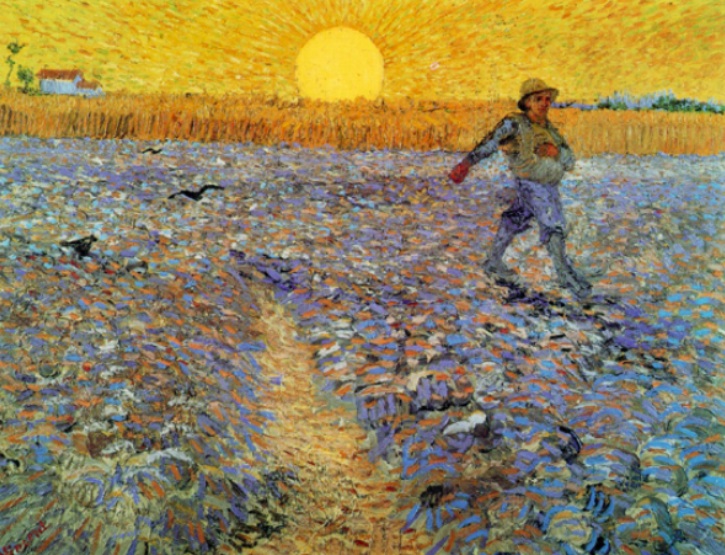}
\caption{The splendid painting  \it The Sower \rm by
Vincent Van Gogh well
 symbolizes the work of
Edoardo Amaldi sower in
Italian postwar physics.}.
\label{seminatore}
\end{figure}

Edoardo Amaldi has been the Sower in
Italian postwar physics, at least until the
70s. With great scientific acumen
 and aware of the responsibility
that events had put him in,
Amaldi launched many seeds on the soil of
Italian physics. Some fell on fertile soil
 and, especially at the University of
Rome,  developed, continuing
the activities initiated by Enrico Fermi with
the group of via Panisperna:  
elementary particle physics, and then the
matter, the physics of universe and gravitation.
A few more attempts, however,
did not catch on completely.

 Amaldi had tried to convince some colleague or student to start an experimental activity  in the   field of General Relativity in Italy. Therefore, when in September 1970 I proposed to him to start an experiment for searching gravitational waves, he was extremely happy, gave all his support and he himself was full time in it. 
 
  The idea to start research in fundamental physics came to me during my stay at the University of Iowa (USA), 
where I had spent a few years doing research on cosmic rays
and Van Allen radiation belts of the Earth. 
Being the assistant of Edoardo Amaldi, during the 
last few years I had heard from him the importance to do experiments in   the new fields of physics: 
gravitational waves (GW) and  the infrared cosmic background. So when I told him, the next day after my 
return from Iowa City, that I wished to start an experiment for the search of 
gravitational waves, his eyes lighted and he stared at me in a way which I 
shall never forget.

\begin {figure}
\includegraphics[width=0.8\linewidth]{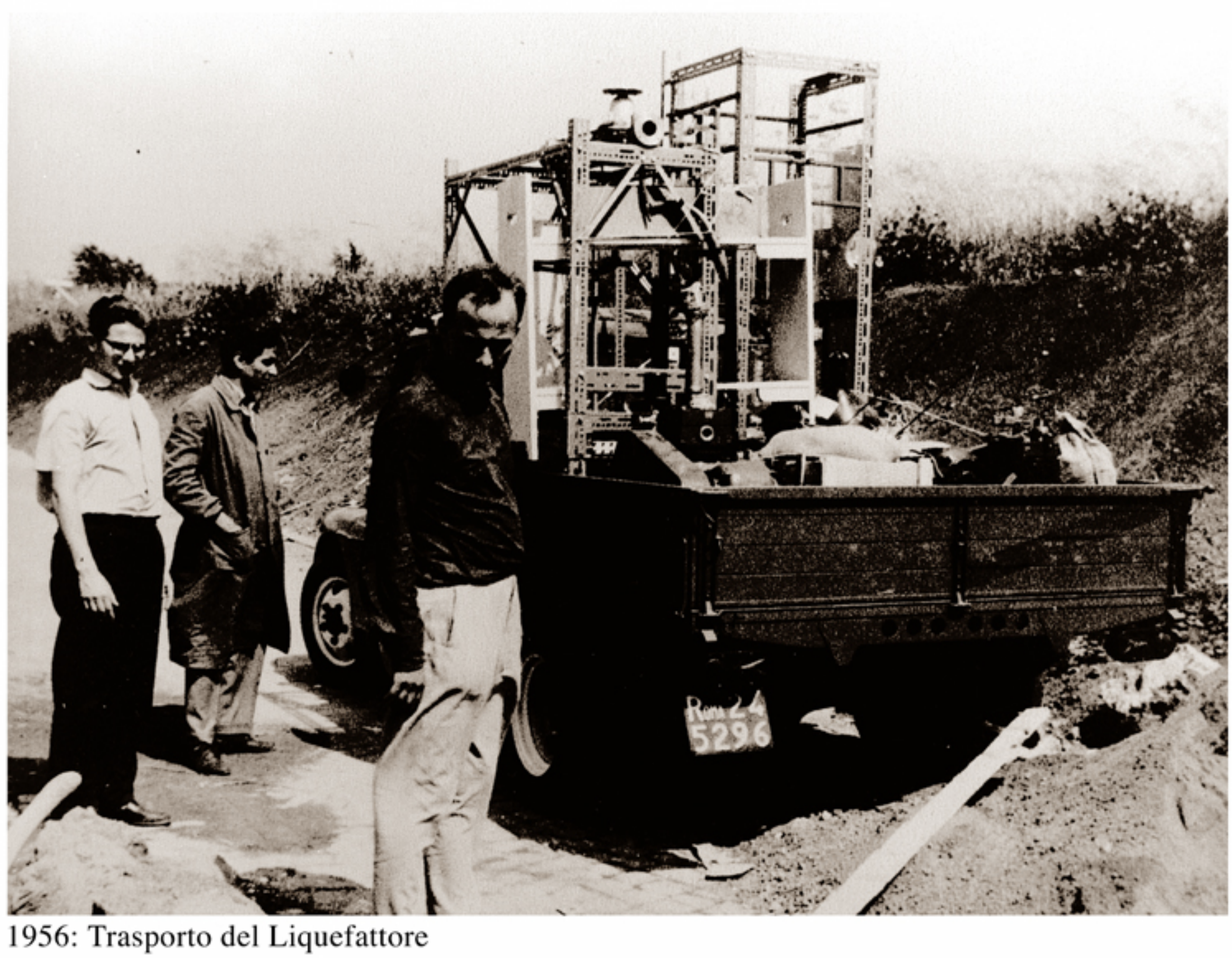}
\caption{The apparatus for the $ He ^ 3 $ in $He ^4 $ } diffusion experiment, built in 1955-1956 at the University of Rome  From left G.Pizzella (INFN felloship), Franco Tesi (truck driver) and J.Reuss (German physicist) \cite{frascati1956}.
\label{frascati1956}
\end{figure}

In January 1971 Remo Ruffini\footnote{Remo Ruffini had just got his \it laurea \rm degree at the University of Rome with a thesis on relativistic astrophysics.}, who was then at the University of Stanford, sent to Amaldi, on a confidential basis, the proposal of William Fairbank (University of Stanford) and William Hamilton (University  of Louisiana) for a large five-ton ultracryogenic  antenna equipped with a  SQUID transducer. Immediately I decided that also in Rome we would have to make a similar experiment. Since we needed a laboratory that could house the antenna and also served cryogenic physicists I proposed Frascati. In Frascati in 1956  I had worked in the installation of the He liquefier and I had performed diffusion experiments $ He ^ 3 $ in $ He ^4 $, first Italian researcher working in Frascati with a scholarship INFN, along with  J. Reuss and with the technicians  Solinas and Bellatreccia, when still the only place fit for use was the laboratory for the helium liquefier (see fig.\ref{frascati1956}).

Amaldi immediately  summoned a meeting with the director of the INFN Laboratories in Frascati, Italo Federico Quercia, who appeared favorable to start this new activity in the LNF. The next day I went to Frascati to discuss it, but it was clear that the interest had been expressed only in words, as there were already more research to be pursued.

However we continued to develop the project, also arousing a lot of interest in theoretical physicists, as Bruno Bertotti, Nicola Cabibbo and Bruno Touschek \cite{tou}. Since the beginning we had the important collaboration of Ivo Modena and Giovanni Vittorio Pallottino\footnote{GianVittorio Pallottino had been an important coworker for my space experiment on the solar wind.}, old fellow adventurers. The experimental activity started at the laboratories of the Snam-Progetti ENI in Monterotondo, by Giorgio Careri who had been the director, for the installation and commissioning of the great detector. SNAM had set up the premises and purchased a liquefier for liquid helium. In the spring of 1974 we moved to Monterotondo where all the pieces began to arrive of the cryostat for the large antenna. I well remember that during this period we had the visit at Monterotondo of Bruno Touschek, with whom we discussed the pilot project ending finally with a toast.
\begin {figure}
\includegraphics[width=1.0\linewidth]{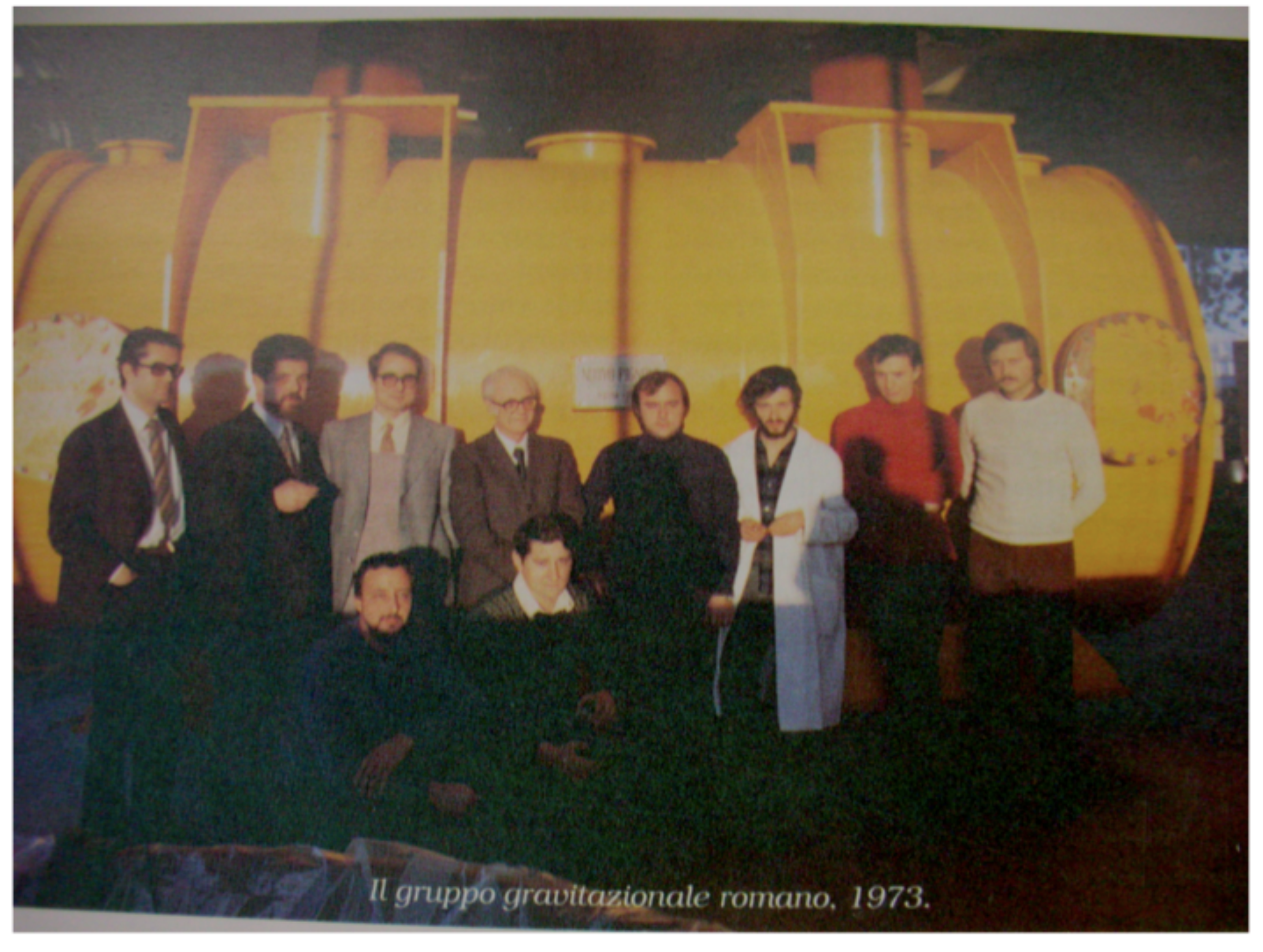}
\caption{EXPLORER at the SNAM-Progetti in Monterorondo. From left: Pallottino, Modena, Pizzella, Amaldi, Serrani, Carelli, Lucano, Giovanardi and, in the lowest line, a technician and Foco.}.
\label{monte}
\end{figure}

In the following years the experiment went on with changing fortunes and, after leaving the SNAM-Projects and after a further attempt to go to Frascati, then unavailable for political reason, we landed in 1980 at CERN, very well received, where finally we realized the cryogenic antenna EXPLORER  cooled to 2 K. This antenna has worked continuously until 2010, when the collaboration with CERN was terminated. As for funding, which until 1980 had been secured by the CNR, they were provided by the INFN.

 \section{Gravitational waves: do they exists ?}
 
 The existence of gravitational waves (GW) was predicted already in 1900 by Lorentz and independently in 1905 by Poincar\'{e}, based on a natural analogy of the  Newtonian field  with the Coulomb field. In 1916 Einstein showed that one of the solutions of its linearized equation of General Relativity   has just the analytical form of a wave, and  showed that it would have been extremely difficult the measurement  due to the very low  interaction with matter. Moreover Levi-Civita, co-founder along with Ricci-Curbastro of the mathematical technique by which Einstein was able to develop the GR, argued that the solution found by Einstein had physical content, being only a wave of mathematical coordinates. I came to know this fact by Edoardo Amaldi, as he told me about the exchange of letters between Einstein and Levi-Civita.
 
 Einstein initially
was convinced that the gravitational waves
 had a reality,
but in 1936 he wrote a paper \cite{eins} with
Nathan Rosen in which, essentially, he resumed
the arguments of Levi-Civita.
The work was rejected by 
Physical Review because it contained an error. However, it is  interesting to  remember
 this story because it indicates that the physical reality
 of the gravitational waves is not
so  certain, and only their experimental revelation
will settle the issue.
 
 Let us briefly recall  the fundamental equation of GR
\be
R_{ik}=\frac{8\pi G}{c^2}(T_{ik}-\frac{1}{2}g_{ik}T)
\label{ricci}
\ee
where $ R_ {ik} is$  the Ricci tensor, $ T_ {ik} $  the energy-momentum tensor, (T  its trace) and $ g_ {ik} $ is  the metric tensor  that comes also in  $ R_ {ik} $ in a nonlinear manner.
The tensor $ g_ {ik} $ is the unknown in Eq. \ref{ricci}  and describes the action of gravity as a deformation of the  space-time geometry.
Linearizing this equation, ie introducing the possibility of weak field $ g_ {ik} = \bf {1} \rm + h_{ik}$ with $h_{ik} <<\bf{1} $ \rm, we get
\be
\Delta h_{ik}-\frac{1}{c^2}\frac{\partial^2 h_{ik}}{\partial t^2} =0
\label{onda}
\ee
in vacuum. Thus the gravitational waves  travel in vacuum with the speed of light. In the following we indicate with $h$ the amplitude of  the GW.

Assuming the existence of GW,   it is possible to calculate the carried power  and it is found that  is extremely small. Already Einstein in 1916 calculated that the power irradiated from any source achievable in a laboratory is so small that \it it has a practically vanishing value. \rm 
For this reason, today only cosmic sources, where huge masses and accelerations are available, are taken into account. Among them we list: \\
 $\bullet $ The GW emitted from binary star systems. Some years ago Hulse and Taylor were awarded the Nobel Prize by measuring the decrease of energy of the binary system PSR 1913+16 and showing, among other things, that it loses energy just as required by GR. \\
$ \bullet $ The GW  emitted by pulsars. This can happen if the pulsar does not have spherical symmetry, so that its tensor of quadrupole varies in time due to the rotation. To give an idea of how small is the signal expected on Earth, if we consider a neutron star with radius of 10 km and with an equatorial asymmetry of $100 ~\mu m $  rotating with a period of $ 1 ~ ms $, we find on Earth, at a distance of 1 kpc, a perturbation of the metric tensor of the order of $ h\sim 5 ~ 10 ^ {- 27} $.\\
$ \bullet $ The GW emitted by supernovae. Also in this case it is necessary that the explosion be non symmetrical. There are many models on the way, but in general we can see that the perturbation of the metric tensor  observed on Earth for a supernova in our Galaxy is of the order of  $h \sim 10 ^ {- 18} $, depending on the model and on the distance.\\
$ \bullet $ The GW emitted by the fall of a star  into a  black hole.\\
$ \bullet $ Finally the GW generated at the Planck time, that is $ 10 ^ {- 43} $ seconds after the big bang. The measurement of such GW should lead information  to understanding the birth of the Universe.

\section{The gravitational waves detectors}

\subsection{Interferometric detectors}

The  arrival of GW in a  region of space  changes the laws of the flat Euclidean geometry,  by introducing a curvature. Imagine a flat plate that, when the wave comes, becomes curved. The consequence is that the distance between two points varies; for example two points at opposite ends  of the plate when this is curved approach. This is the principle on the basis of operating interferometers.

The interferometer consists of a laser light from a point that goes, through a beam splitter, in two directions perpendicular to each other, it is reflected by two mirrors placed at the same distance (3 km in the case of VIRGO) and returns to the starting point. Upon arrival the two beams are no longer in phase, as they were at the start, because  one of the two beams has traveled a different distance from the other.

\begin{figure}[ht!]
\centering
\includegraphics[width=120mm]{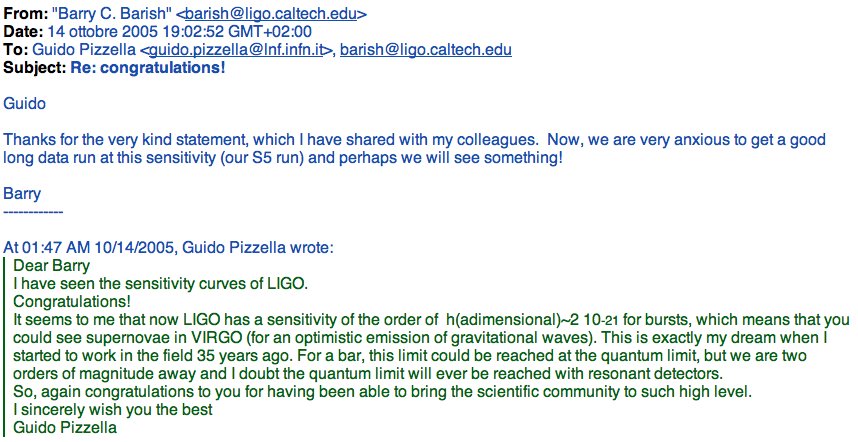}
\caption{ Exchange of letters  with the former LIGO principal investigator Barry C.  Barish.}
 \label{barish}
\end{figure}

So far  two LIGO interferometers (arms 4 km) installed in Livingstone in Louisiana and Hanford in Washington State, VIRGO (arms 3 km) installed in Cascina and GEO (arms 600 m) installed in Hannover in Germany have been from time to time in operation. The most ambitious  one is certainly the LIGO project, which started a few decades ago and reached a sensitivity greater than that of the bar detectors in 2005, as shown in an exchange of letters  with the former principal investigator Barry C.  Barish (see fig. \ref{barish}).

About VIRGO, I think it was an error by the European scientific community not to launch two interferometers instead than just one. This is because the search for such feeble signals needs the use of a two, at least, coincidence experiment (see later  the section \it search for coincidences\rm). This is what they do in USA with two LIGO and, if GW will be discovered \cite{scope}, the credit shall not belong to the European Science.

\subsection{The resonant detectors}
Joe Weber of the University of Maryland had, in the 50s, a very ingenious idea. He thought  that the perturbation $ h $ of the metric tensor would have put in vibration a massive bar. To get an idea of the magnitude of these vibrations you can  consider the explosion of a supernova in our Galaxy (of course, rule out the chance that the explosion takes place in the vicinity of the Earth, because if this would happen, life on Earth would end). It is estimated that a galactic supernova would vibrate the bar with an amplitude of the order of $ 10 ^ {- 18} ~ m $. This vibration is faced with the vibrations caused by the thermal noise of the bar. A bar of mass $ M $ = 1000 kg cooled to the temperature of 1 K has a thermal vibrations on the order of $ 10 ^ {- 17} ~ m $,  greater than the signal we want to measure. To this noise  an equal noise due to electronic devices shall be added.

From here we understand the great difficulty of the experiment that seeks signals due to GW, as  the gravitational forces generating the signal compare to the much more large electromagnetic forces which produce the noise. The problem is faced by cooling the detector and using optimized filtering algorithms  in order to extract small signals in the presence of a large noise.

Despite these difficulties various experimental groups in the world decided to take the road to study such a fleeting phenomenon. Some groups terminated their activity very soon, among them a group in Canada ed one in Rochester.
The group at Stanford, founder of the cryogenic  detectors, decided to end the experiment when, on October 17th, 1989, an earthquake badly  hurt their cryogenic antenna. Five groups succeeded and begun to put into operation cryogenic detectors: 
in 1990 EXPLORER at CERN, in 1991 ALLEGRO  in Louisana, in 1993 NIOBE in Australia, in 1994 NAUTILUS in Frascati and in 1997 AURIGA in Padua, this last one a replication of NAUTILUS. 

Except for NAUTILUS and AURIGA, the operation of the resonant detectors has been terminated, because the laser detectors, though not able to reveal massive particles, are much more sensitive for the detection of gravitational waves. The last two resonant detectors NAUTILUS and AURIGA will be turned off in June 2016\footnote{These detectors were supposed to be turned off at the end of 2015. Their life  has been extended to cover a few months in 2016, because, during this period, the operation of the interferometers is stopped for the work needed to improve their sensitivity.}

\subsection{The EXPLORER resonant detector}

The gravitational wave detector EXPLORER consists of a bar of aluminum 3 meters long, with a diameter of 60 cm and a mass of 2270 kg. Upon arrival of the gravitational wave it should vibrate at its longitudinal resonance frequency  $ \nu = \sim915 $  Hz,  with amplitude  extremely small, order of $h$. The vibration is detected by means of a capacitive electromechanical transducer, consisting of a capacitor, one plate of which is fixed and the other one  vibrates at the same frequency of the bar, a system of two coupled oscillators. The distance between the plates varies when the bar is solicited by a GW or by noise. An electric charge is put on the capacitor and generates a signal of variable voltage when the distance between the plates varies due to the vibrations. The signal is amplified by a SQUID \footnote {The SQUID are superconducting devices which measure weak magnetic fluxes, in our case obtained by sending the signal from the transducer in a reel. An important use of these devices is the study of weak electrical currents generated in the biological brain circuits. It was also attempted to apply them in cybernetic circuits, so far without a complete success.} and recorded.

The transducer was built in the laboratories of Frascati CNEN (hereinafter ENEA) under the direction of Roberto Habel. It is important to remember this because it shows a first important  collaboration  for the search of gravitational waves with a physics group in Frascati.

 A fundamental step to take when a new instrument is built is its calibration.
For Explorer we have used  a time-varying gravitational field generated by a rotor with frequency half of the resonant frequency of the bar \cite{rotore}. The measurements were in perfect agreement with the expected values  and this gave us confidence in the reliability of our instrumentation.

Given the smallness of the signal expected for the GW we must take many precautions. First thermal noise  must be made the smallest possible. This is done by putting the bar in a cryostat, that is in a container cooled with liquid helium (4.2 K at atmospheric pressure). The cryostat, consisting of several cylindrical containers each of which at a temperature decreasing towards the interior, must be as more as possible isolated from external mechanical disturbances. This is obtained by suspending the various cryogenic vessel of the cryostat by means of cables which play the role of mechanical filters and finally by suspending the bar in the most interior container  with a cable that wraps partially below  its gravity center section. In this way we get a good mechanical attenuation in total of approximately 200 dB, which assures the attenuation of mechanical disturbances\footnote {The detector moreover is sensitive to even minor earthquakes, but this is taken into account  by the help of seismographs and especially by the coincidences with another detector located at a great distance.}.

EXPLORER  was the  first cryogenic detector to operate continuously since the year 1990 at temperature of 2 K with good sensitivity, the greatest sensitivity until the year 2000.

\section{Going back to the Frascati Laboratories: NAUTILUS}

On December 5, 1989 suddenly Edoardo Amaldi, the Sower of the Italian Physics, passed away.  A few days later, on December 18, I was in my office at the University \it La Sapienza \rm in Rome, when I received a phone call from Enzo Iarocci (...  tip in a dream by Edoardo Amaldi? ...). Enzo, who was to become director of the LNF from January 1, 1990, was proposing to bring the NAUTILUS detector under construction at CERN in the Laboratories of Frascati, were liquid helium liquefiers were available. We recall that NAUTILUS was under construction at CERN because this was the only lab available for our group, but the idea was to carry NAUTILUS in another place as soon as the construction was over, in order to be able to study events in coincidence between two detectors installed in far away places.

\begin{figure}
\includegraphics[width=1.0\linewidth]{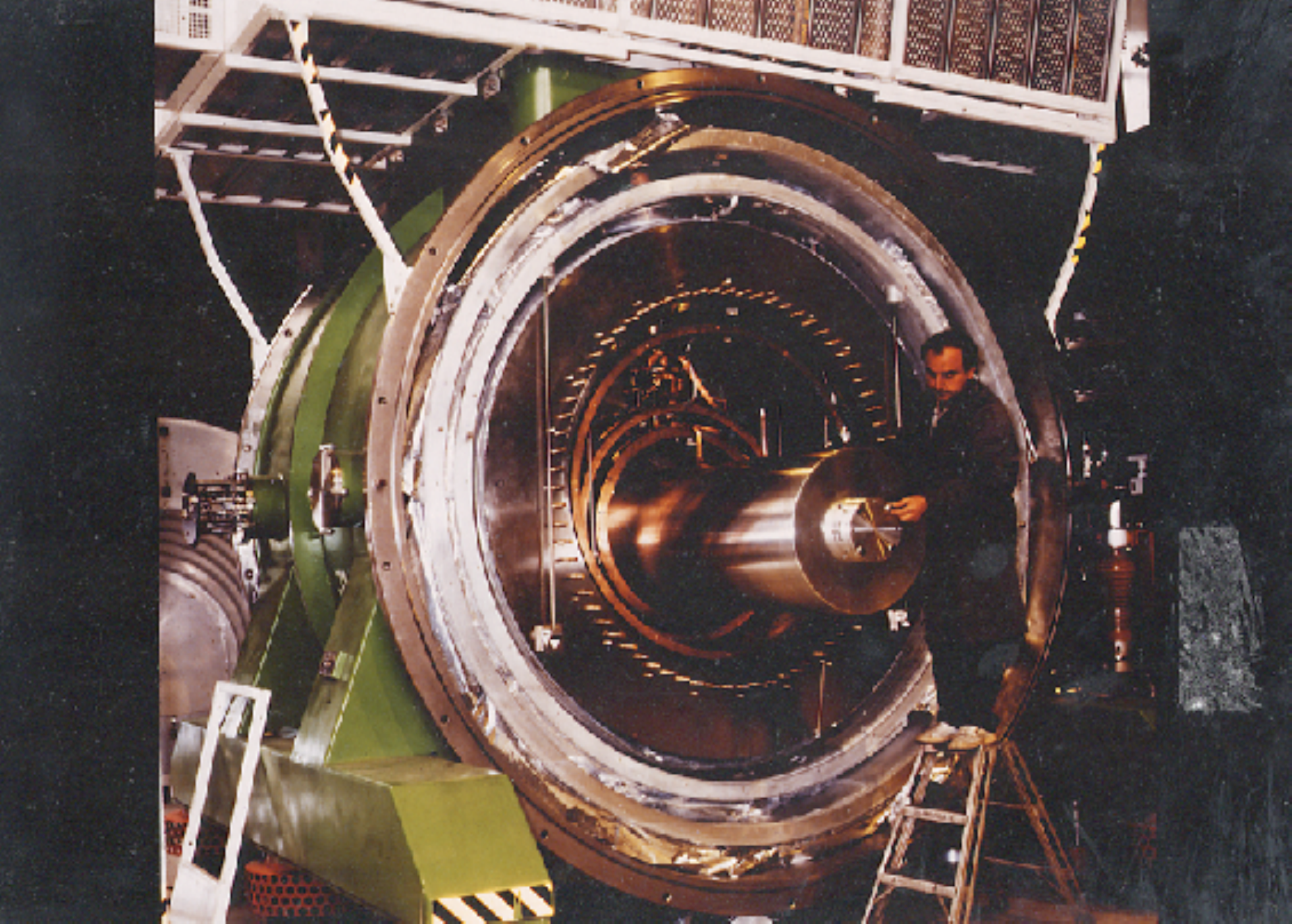}
 \caption{
 The GW detectori NAUTILUS at the  INFN Laboratories in  Frascati.
        \label{nautilus} }
\end{figure}

\begin{figure}
\includegraphics[width=1.0\linewidth]{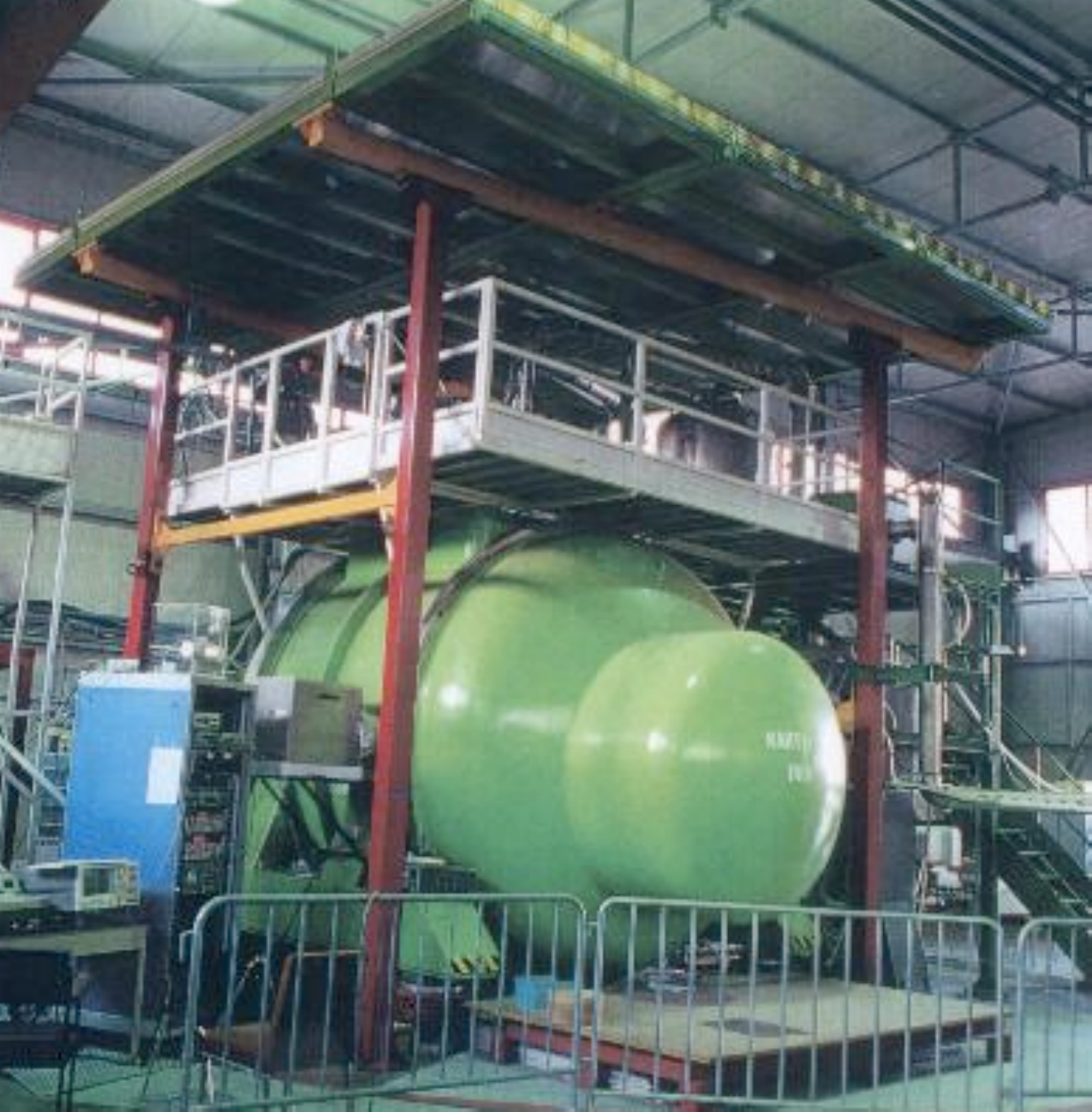}
 \caption{
 NAUTILUS with the cosmic ray detectors.
        \label{nautilusr} }
\end{figure}

 Soon after we met in the LNF.  Iarocci showed me the building where to put the NAUTILUS. In about a year the transfer was made. Before we  had taken steps to build a metal swivel where NAUTILUS  would have been installed. NAUTILUS is shown in figures \ref {nautilus} and \ref{nautilusr}. NAUTILUS,  with a bar identical to EXPLORER, has been equipped with a dilution refrigerator which allows a cooling of the bar down to 0.1 K in order to increase the sensitivity  by reducing the thermal noise. This temperature was never reached so far for large bodies, so that we could say that NAUTILUS is the coldest heavy body in the Universe, unless you consider the possible existence of other intelligent beings.

\section{Search for coincidences}

The major problem in all the GW experiments is the smallness of the
signals (due to the gravitational force) compared with the noise
(due to electromagnetic forces). Thus we are forced to operate in
conditions of very small signal-to-noise ratios. In this situation,
even in presence of well behaved noise, we must take into consideration
as GW candidates a large number of data, in the very great majority
due to noise but which could embed rare and precious information
on possible GW. Neither is possible to reduce significantly the
number of candidate data by a careful screening, using information
from different instrumentation.

For example, suppose the detector has a well behaved noise expressed in unit of kelvin, $T_{eff}=5~mK$
with a Maxwellian distribution, and a bandwidth of 10 Hz. 
In one hundred days we have $100\cdot864000$
independent samples and the number of
samples with energy, say, $E>50~mK$ (that is $SNR\ge10$) is
\be
N(E>50 mK)=100\cdot 864000e^{-\frac{50}{5}}\sim3900
\ee
a large number.

In the real case, with additional noise of unknown origin, we may have
a number of samples above threshold one or two orders of magnitude greater.
Eliminating  the data correlated with seismic
signals, as done by the Rome group,  the number of data reduces  by only a few percent.

It is possible, however, to improve considerable the search of GW by
employing the coincidence method, as initially done by Weber. For example,
suppose we have two equal detectors, each detector providing 10000
candidate data (signals above a given threshold) over a period of one hundred days. With a coincidence window of 0.1 s ($\pm0.05~s$) we have
an average number of accidentals
\be
\bar{n}=\frac{10^4~10^4~0.1~s}{100~86400}\sim 1
\ee
that is, we reduce by a factor of ten thousand the candidate data to
consider as possible GW.

This argument appears rather obvious:\\
\it give most importance to the coincidence technique  for
cleaning the data, much more than any other technique.\rm \\
Yet many people do not give proper consideration to that. I have had several discussion with other scientists, who treat the data before searching for coincidences  on the basis of theoretical expectations and do not care for another detector to compare with. As example, the VIRGO experiment, just one detector, initially designed to detect waves possibly generated by pulsar 

In my opinion the analysis  of data should consist essentially
in comparing the  signals in time coincidence among two or more detectors. In the case of coincidence search the background can be obtained, for example, by shifting several times the time of occurrence of the events  of one of
the  detectors\footnote{It can be shown that this experimental background determination  has
to be preferred to the random reshuffling procedure.}.

With this experimental procedure 
one circumvent also the problems arising from a non stationary
distribution of the events.

A different reliable approach is  provided by the Bayes theory.
According to this theory the probability to have a certain result
depends on the degree of belief, due to
previous information, and on the statistical computations with the new data.

\subsection{Search for coincidences with EXPLORER and NAUTILUS}
 
We have applied  the Bayes method for determining upper limits to the GW searched with  resonant detectors. The result and the procedure has been described in  ref.\cite{bayes}.

Using the most common procedure for the search of coincidences we have processed the data obtained with NAUTILUS  in time coincidence with other operating resonant detectors, particularly EXPLORER. In the following we describe very briefly some results. 

The most interesting one has been obtained with data recorded during 1998
in the period 7-17 September when an intense
activity of the black hole candidate XTE J1550-564, and within days of the giant flare
from the magnetar SGR1900+14 occurred. During this period we had 21 coincidences ($\pm 0.5$ s) between signals recorded by the two detectors, while expecting on average only 8.1 due to chance. Furthermore there was also a  significant correlation  between the coincidences and the onset of the X-ray emissions \cite{modena}. One coincidence was due to the largest events ever found, in coincidence, in all our experiments, 5.7K for EXPLORER and 5.8K for NAUTILUS. It is unfortunate that no other GW detectors was  in good operation during 1998. In my opinion,  we missed the best opportunity for an important discovery. In the rest of the year 1998  over a total period of 95 days we found 61 coincidences while expecting by chance 50 \cite{c1998}.

Other intriguing  results were obtained  in the year 2001 with
 data collected by EXPLORER and NAUTILUS, for a total measuring time of 90 days \cite{c2001}. We repeated the coincidence search as with the 1998 data, using the same  algorithms based on known physical characteristics of the detectors. An interesting feature was found  during the sidereal hours in the range 2-5, when the detectors were favorably oriented with respect to the Galactic Disk. We found 8 coincidences while expecting, by chance, 3.1.
 
 Similar result was also obtained with data recorded during 2003. In the range 2-5 sidereal hours we found 6 coincidences while expecting, by chance, 1.6.
 
 Such coincidence excesses were not found after the year 2003.
 
 It is evident that these results,  although  intriguing, are not sufficient to make any claim, taking also into account that no other GW detector  supported them. The real problem with the resonant detectors is their  poor sensitivity, unless we believe that cooperative mechanisms are operating during the interaction of GW with the detectors, as suggested by Giuliano Preparata \cite{prepa}.

 \subsection{Signals from cosmic rays}
 
 Contrary to the interferometric detectors, the resonant detectors are sensitive to the passage of particles.
 We have used this feature especially to perform a sort of \it absolute \rm calibration of the bar-detector, measuring the relationship between the bar signal and the energy deposited in the bar, therefore we are sure that the tiny signals seen by the detector, extracted by means of optimized algorithms, correspond to  definite amounts of released energy. No other GW experiment has done it, and the usually adopted calibration procedure requires a modeling of the calibration apparatus.

Ê The signals are due to the mechanical vibrations produced by the expansion that has along the path of the particles  because of the warming for to energy dissipation.  Therefore the signals depend on the ratio of the thermal expansion coefficient to the specific heat, that is the Gr$\ddot{u}$neisen. coefficient. This is independent of temperature at least until  the material becomes superconducting (1 K for aluminum). The various acoustic models give the energy $ \epsilon $ expected in resonant mode detectors
Ê \be
\epsilon = 7.64 \cdot 10^{- 9} ~ W^2 \cdot ~ f
\label {Kgev}
\ee
where $\epsilon $ is expressed in kelvin, W (in GeV) is the energy dissipated in the bar, and $ f $ is a geometric factor of the order of  unity.

\begin {figure}
\includegraphics[width=0.8\linewidth]{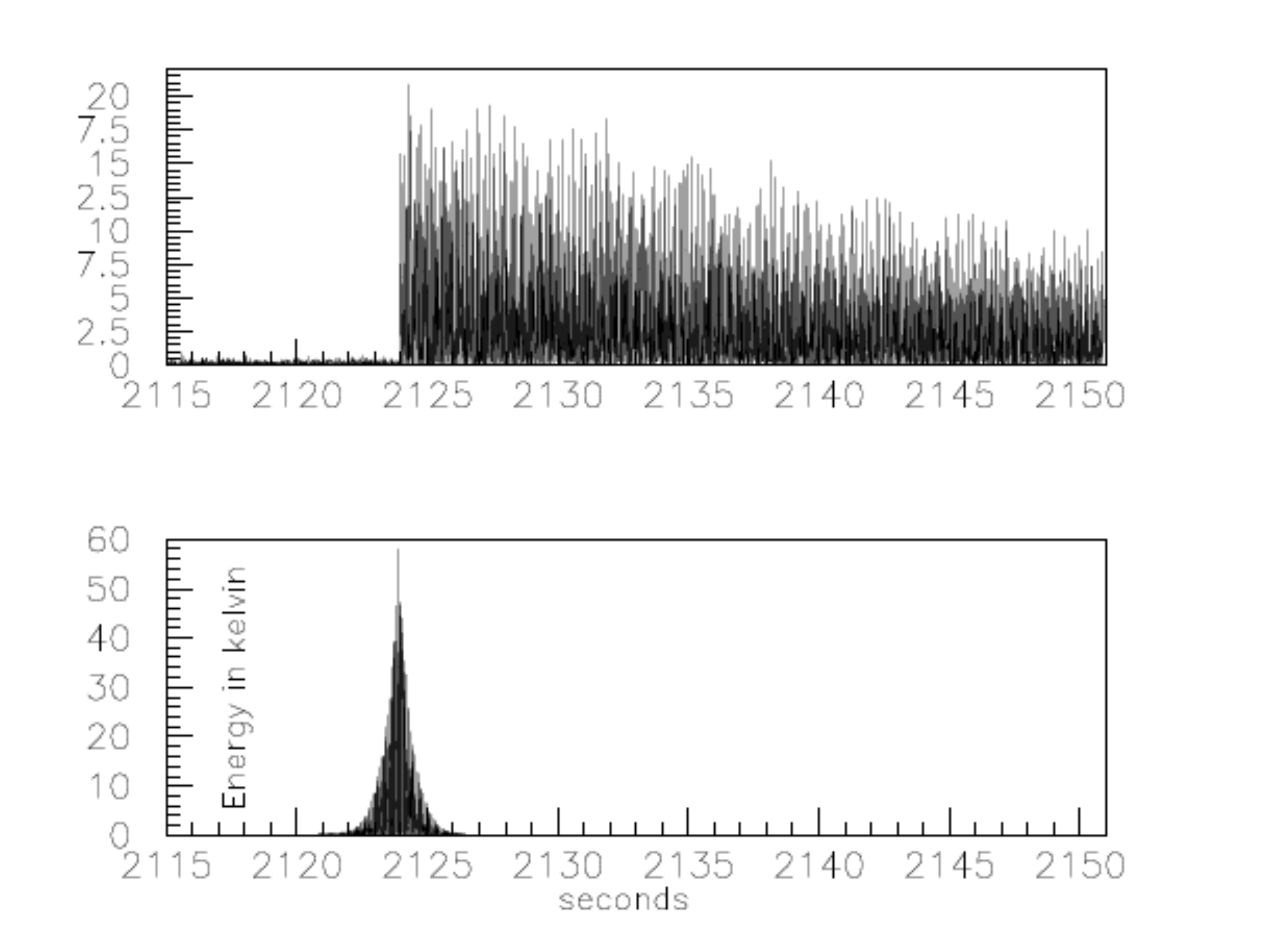}
\caption {
The signal (volt square) before optimum filtering versus the UT. The time is
expressed in seconds, from the preceding midnight. From the decay we
evaluate the merit factor of the apparatus, $Q = 1.7 \cdot 10^5$. The lower figure shows
the data after filtering, in unit of kelvin. Here the reverse of the decay time
gives the detector bandwidth equal to 0:34 Hz .}
\label {bigone}
\end {figure}

Measurements were performed with NAUTILUS  and EXPLORER  using equipment for cosmic rays designed and built by Francesco Ronga. One of the results \cite{bigone} is shown in Figure \ref {bigone} where a very large signal due to a cosmic ray shower is reported.

We also  found signals larger than calculated with Eq. \ref {Kgev}   when the detector was cooled to 0.1 K,.

In order to obtain experimentally the relationship between the energy of the particles interacting with the detector and the generated signal, we performed  an experiment, called RAP at the INFN Laboratories using a small aluminum rod cooled to 0.1 kelvin and subjected to the electron beam  from DAPHNE.

\begin {figure}
\includegraphics[width=1.0\linewidth]{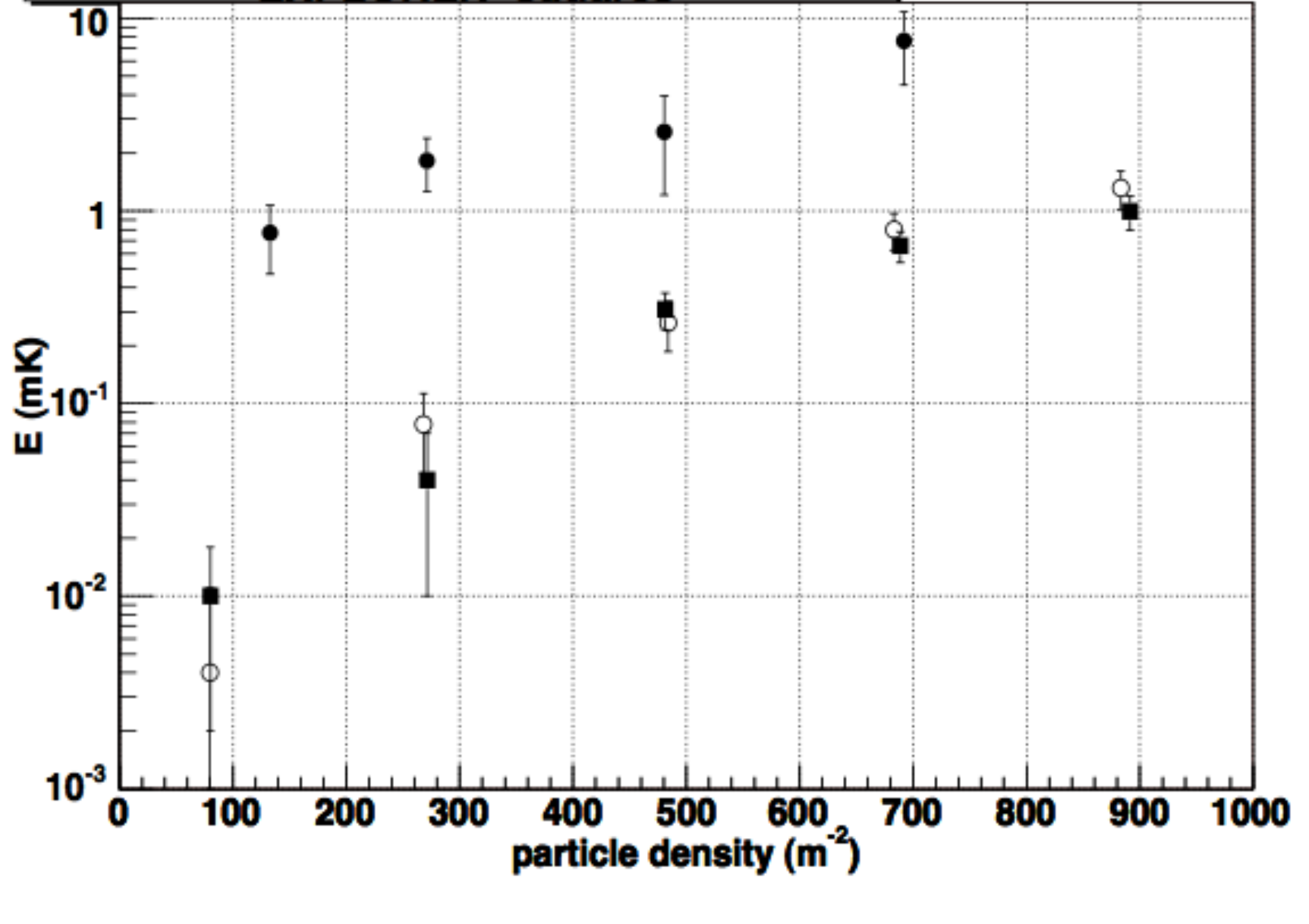}
\caption{Filled circles NAUTILUS at T = 0.14 K, open circles NAUTILUS at T = 3 K,
Filled squares EXPLORER at T = 3 K. The data gathered at T = 0.14 K are roughly one
order of magnitude larger than those collected at T = 3 K.}
\label{rap}
\end{figure}

The result \cite{grune} is shown in Fig. \ref {rap}.  We found an increase of about one order of magnitude in the energy of the signals,  due to the superconducting state of the aluminum.

 As a by-product NAUTILUS has led interesting contributions in setting upper limits on exotic components of cosmic radiation \cite{quark}.
This search has been carried out using data from NAUTILUS and EXPLORER equipped with cosmic ray shower detectors. We remark that the particle detection mechanism is completely different and  more straightforward than in other cosmic ray detectors. The results of ten years of data from NAUTILUS (2003-2012) and 7 years from EXPLORER (2003-2009), searching  nuclearites of mass less than $10^{-4}$ gram, show a flux smaller than predicted considering nuclearites as dark matter candidates.

\section{Final consideration}

\begin{figure}[ht!]
\centering
\includegraphics[width=130mm]{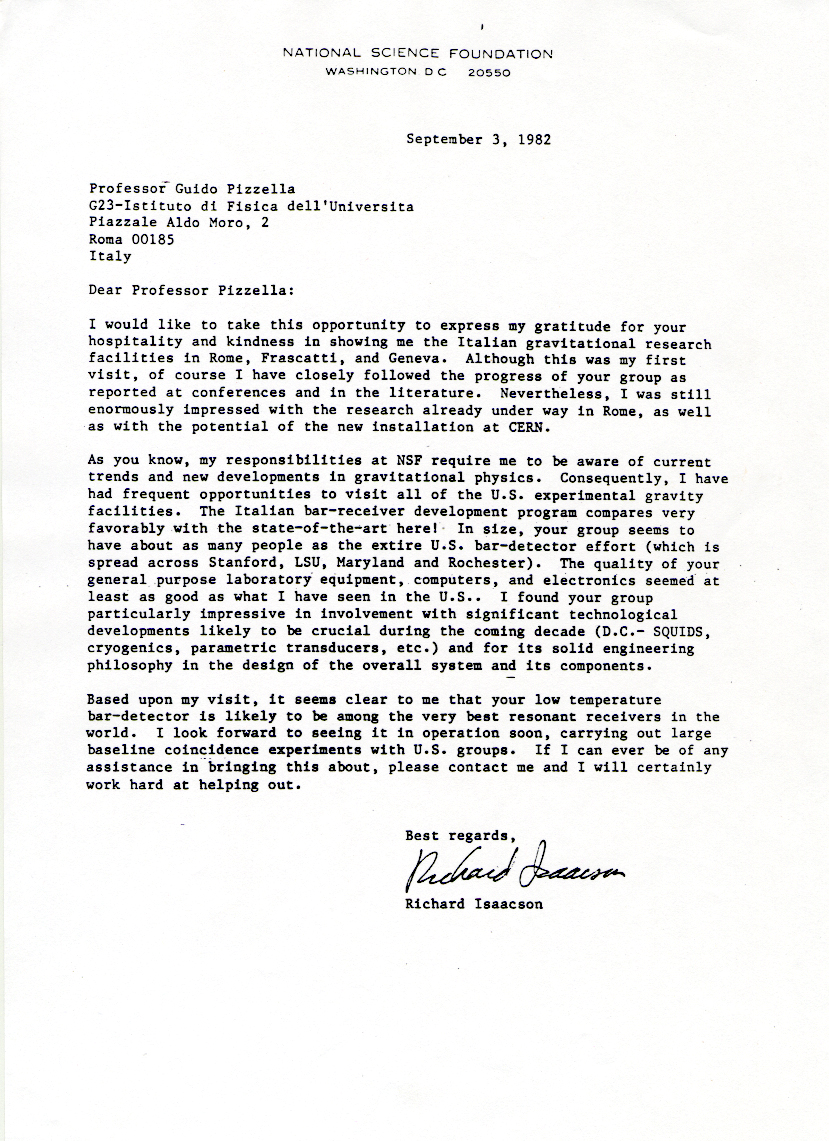}
\caption{Letter by Isaacson in 1982. This marks the date when the project LIGO  has received a boost. }
 \label{lsa}
\end{figure}

As recognized also by members of the American NSF, as shown in fig.\ref{lsa},
Êthe search for gravitational waves carried out by  the Rome group\footnote {Over the years the activities carried on  by the Rome group of gravitational waves relied on contributions from several researchers listed here: Edoardo Amaldi, Pia Astone, Danilo Babusci, Massimo Bassan, Romano Bizzarri, Paolo Bonifazi, Franco Bordoni, Pasquale Carelli, Gabriella Castellano, Giorgio Cavallari, Eugenio Coccia, Carlo Cosmelli, Sabrina D'Antonio, Antonio Degasperis, Viviana Fafone, Valeria Ferrari, Sergio Frasca, Franco Fuligni, Gianfranco Giordano, Umberto Giovanardi, Roberto Habel, Valerio Iafolla, Ettore Majorana, Alessandro Marini,  Evan Mauceli, Yuri Minenkov, Ivo Modena, Giuseppina Modestino, Arturo Moleti, GianPaolo Murtas, Yujiro.Ogawa, Gianvittorio Pallottine, Guido Pizzella, Lina Quintieri, Piero Rapagnani, Fulvio Ricci Alessio Rocchi, Francesco Ronga, Roberto Terenzi, Guido Torrioli, Massimo Visco, Lucia Votano. It  has been their dedication to this difficult experiment that allowed the group to position itself at the forefront in the world of search for gravitational waves.} played an important role in the scientific landscape. The international scientific community  has given the name \it Edoardo Amaldi \rm to the most  important Conference, repeated every two years in different countries, for the search for gravitational waves.
 \begin {figure}
\includegraphics[width=0.8\linewidth]{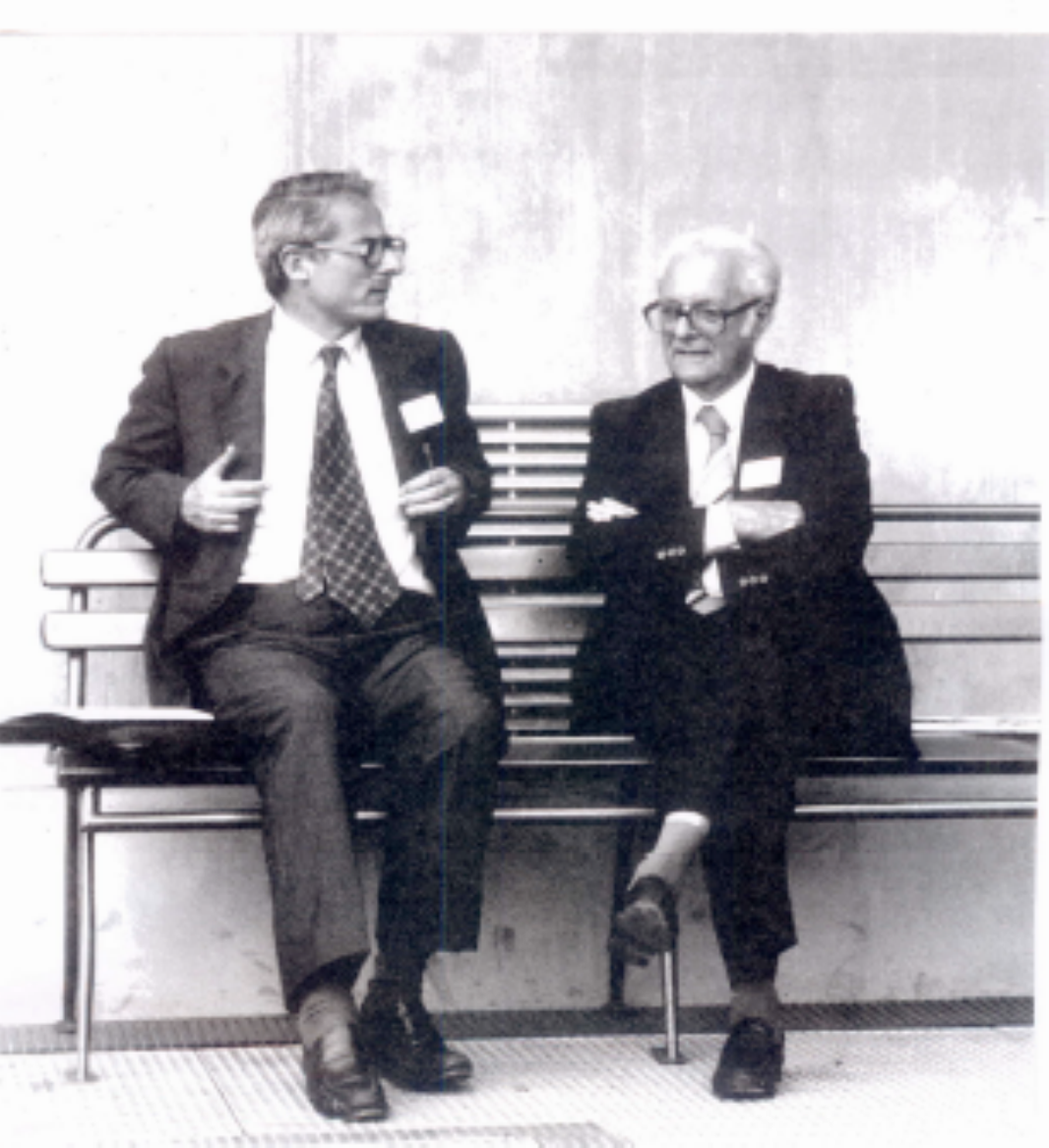}
\caption{ Edoardo Amaldi and Guido Pizzella at the General Relativity Conference
in Padua in 1983. Photo shot by Emilio Segre.}.
\label{amaldipizzella}
\end{figure}

 Although the discovery of gravitational waves has remained a mirage for us, I believe we have brought  contributions to the development of this research, culminating in the large interferometers \cite{scope}, and in the space experiment LISA, full of hope, which soon will tell us whether the mirage reflects a new reality.

Everything  was made possible by the vision of Edoardo Amaldi, both for his contribution to the creation of the National Laboratories of Frascati and CERN, both for his open to any new scientific enterprise. To this day it is unlikely to find such Builders of Science. The succession of life leads to the Big Science with thousands of researchers who trooped, and  hardly they can see   other possible horizons.
  
 An exception to this general trend  when accidentally found myself in September  2009 alongside Marcello Piccolo, during a fifty-year marriage celebration of our common friend Lina Barbaro Galtieri.  I proposed to Marcello a  new experiment, to be carried out in Frascati, for the measurement of the propagation of the Coulomb electric fields. The idea of this experiment had come by considerations  on the physics of systems of reference and therefore I regard it a natural consequence  of research in the field of General Relativity.

Marcello, while engaged in other experiments, agreed. The experiment, which aroused also the interest of Giorgio Salvini\footnote{Salvini [1921-2015] had been the first director of the Frascati National Laboratories.}, gave the result described in the paper \cite{marcello}.
With this experiment we found that a moving electric charge carries  \bf rigidly \rm its own  Coulomb field. The scientific significance of this result  has not  yet been evaluated in full.

We are now trying to repeat this experiment at the Frascati National Laboratories (sixty years after my first experiment here) or elsewhere. Despite some of my previous considerations, I remain very optimistic about the future of Science.

\section{ACKNOWLEDGMENT} 
I thank my wife Elena for her continued encouragement.
This paper is part of an activity aimed to give a contribution to the history of the INFN Frascati National Laboratory.  
I am grateful in particular to Lia Pancheri, for asking me to give a memory about the search of gravitational waves at LNF. I thank also Giorgio Capon,  Giovanni Vittorio Pallottino and Francesco Ronga for useful discussions and suggestions.


\begin{thebibliography}{99}

\bibitem{frascati1956}http://www.lnf.infn.it/lnfimages/ , area: historical , liquefattore
\bibitem{tou}Bruno Toushek  had a very  important role in the experiments performed at LNF, with great impact on Physics all over the world. \\
a) E. Amaldi.  1981, The Bruno Touschek Legacy,  CERN 81-19, available at http://cdsweb.cern.ch/record/135949/files/CERN-81-19.pdf\\
b)  Luisa Bonolis, Giulia Pancheri, \it  Bruno Touschek: Particle physicist and father of the e+e collider\rm
, Eur.Phys.J. H36 (2011) 1
e-Print: arXiv:1103.2727.
\bibitem{eins}\it Controversies in the history of the radiation reaction problem in general relativity \rm \\ 
Daniel Kennefick. Apr 1997. 33 pp. 
e-Print: gr-qc/9704002 | PDF
\bibitem{scope} Recently there have been rumors that the two LIGO have detected one signal satisfying the requirements to be due to  gravitational waves.\\
 Nature, 30 September 2015.
\bibitem{weber} J. Weber, Phys. Rev. Lett. 22, 1320 (1969)
\bibitem{bayes} \it On upper limits for gravitational radiation \rm, 
P. Astone , G. Pizzella , Astropart.Phys. 16 (2002) 441-450 

\bibitem{modena}a) \it Coincidences between the gravitational wave detectors EXPLORER and NAUTILUS in 1998, during the activities of the black hole candidate XTE J1550-564 and the magnetar SGR1900+14 \rm 
I. Modena, G. Pizzella, 
 Int.J.Mod.Phys. D15 (2006) 485-491 \\
b) \it Studying the coincidence excess between EXPLORER and NAUTILUS during 1998 \rm , 
D. Babusci, G. Cavallari, G. Giordano, I. Modena, G. Modestino, A. Moleti, G. Pizzella 
 May 2005, arXiv:astro-ph/0505600.
\bibitem{c1998} P. Astone et al., \it Study of coincidences between resonant gravitational wave detectors \rm
Class.Quant.Grav. 18 (2001) 243-252


\bibitem{c2001}
 \it Study of the coincidences between the gravitational wave detectors EXPLORER and NAUTILUS in 2001 \rm 
P. Astone et al.
Class.Quant.Grav.19:5449-5463,2002
 \bibitem{prepa}G.Preparata, Modern Physics Lett. A, 5 (1), 1-5, (1990)
 \bibitem{rotore} Astone P. et al., \it Evaluation and preliminary measurement
of the interaction of a dynamical gravitational near field
with a cryogenic gravitational wave antenna \rm,
  Zeit. Script C - Particles and Fields 50, 21-29 (1991)
\bibitem{bigone}\it Energetic cosmic rays observed by the resonant gravitational wave detector NAUTILUS \rm \\ 
P. Astone et al.. Sep 2000. 8 pp. 
Published in Phys.Lett. B499 (2001) 16-22 
\bibitem{grune} \it Vibrational excitation induced by electron beam and cosmic rays in normal and superconductive aluminum bars, \rm
M. Bassan  et al..  Nucl. Instrum. Meth. A659 (2011) 289-298.
\bibitem{quark}\it Quark nuggets search using 2350 Kg gravitational waves aluminum bar detectors \rm  
P. Astone  et al.. Jun 21, 2013. 
Conference: C13-07-02 Proceedings 
e-Print: arXiv:1306.5164 [astro-ph.HE] | PDF

\bibitem{marcello} \it  Measuring propagation speed of Coulomb fields \rm,
R. de Sangro et al., Published on Eur. Phys. J. C (2015) 75: 137.

\end{thebibliography}
\end{document}